\documentclass[%
 reprint,
amsmath,amssymb,aps,
]{revtex4-1}

\usepackage{lipsum}
\usepackage{graphicx}
\usepackage{dcolumn}
\usepackage{bm}
\usepackage{natbib}

\usepackage{color}
\usepackage{xspace}
\usepackage{amsmath}
\DeclareMathOperator{\Tr}{Tr}

\def\rhoD{\rho_{\Sigma}}
\def\ND{N_{\Sigma}}
\newcommand{\fig}{FIG. }

\def\DE{D_\text{E}}
\def\Dr{D_\text{r}}
\def\DA{D_\text{A}}
\newcommand{\ie}{\textit{i.e.}\xspace}

\newcommand{\etal}{et al.\xspace}

\def\rhoprod{\rho_{\pm}^2}

\begin{document}


\title{Insensitivity of active nematic dynamics to topological constraints}

\author{Michael M. Norton}
\author{Arvind Baskaran}
\author{Achini Opathalage}
\author{Blake Langeslay}
\author{\\Seth Fraden}
\author{Aparna Baskaran}
\email{aparna@brandeis.edu}
\author{Michael F. Hagan}
 \email{hagan@brandeis.edu}

\affiliation{%
Physics Department, Brandeis University, Waltham, Massachusetts 02453
}

\date{\today}

\begin{abstract}
Confining a liquid crystal imposes topological constraints on the orientational order, allowing global control of equilibrium systems by manipulation of anchoring boundary conditions. In this article, we investigate whether a similar strategy allows control of active liquid crystals. We study a hydrodynamic model of an extensile active nematic confined in containers, with different anchoring conditions that impose different net topological charges on the nematic director.  We show that the dynamics are controlled by a complex interplay between topological defects in the director and their induced vortical flows.  We find three distinct states by varying confinement and the strength of the active stress: a topologically minimal state, a circulating defect state, and a turbulent state. In contrast to equilibrium systems, we find that anchoring conditions are screened by the active flow, preserving system behavior across different topological constraints. This observation identifies a fundamental difference between active and equilibrium materials.
\end{abstract}

\pacs{Valid PACS appear here}

\maketitle

\section{\label{sec:level1} Introduction}
Boundary conditions and topological constraints enable long-ranged control over the structural order of equilibrium (passive) liquid crystal systems. This understanding has led to numerous practical applications, most notably liquid crystal display devices and more recently self-assembly of colloids \cite{Bisoyi2011,Senyuk2013, Luo2016, Peng2016}. A similar potential should exist in \emph{active} liquid crystal systems, which are collections of rodlike particles continuously driven away from equilibrium by energy input at the scale of the particles \cite{Simha2002,Saintillan2013a,Marchetti2013}. Indeed, experiments and theory have shown that introducing boundaries into active systems can generate system-spanning effects \cite{Voituriez2005, Edwards2009, Giomi2011, Giomi2012, Woodhouse2012, Wioland2013,Ravnik2013, Doostmohammadi2016, Wioland2016, Shendruk2017,Gao2017, Wu2017, Doostmohammadi2017}. However, in contrast to equilibrium materials, the constituent units of an active material generate hydrodynamic flows that can couple to or compete with the structural order and topological constraints imposed by a boundary. It is unclear how the interplay between flow and boundary-imposed order controls the emergent spatiotemporal behaviors of active materials. This limitation prevents rational design of active devices that might be used, for example, to extract work \cite{Sokolov2010, Thampi2016} or drive assembly.

In this article we theoretically study the interplay between the topological and hydrodynamic aspects of confinement on a class of active materials, extensile active nematics. While previous numerical studies of confined active nematics have led to important insights, \cite{Woodhouse2012, Doostmohammadi2016, Gao2017, Shendruk2017, Zhang2016}, the dependence of their dynamics on container boundary conditions has not yet been studied. Here, we investigate active nematics under circular confinement in containers with four different anchoring conditions, which lead to three different topological constraints on the enclosed nematic director.

Remarkably, in contrast to the case of passive nematics, we find that topological constraints weakly impact the structure of active flows. In all containers, the interplay between topological defects, their self-generated flows and boundary constraints leads to a rich, but similar, set of spatio-temporal dynamics. As confinement is increased or active stress strength decreased, the system transitions from a turbulent state to a static configuration resembling a confined passive nematic. In between, the system exhibits a unique dynamical steady state characterized by a pair of co-rotating $+\frac{1}{2}$ defects which undergo spontaneous and continuous flow, with $-\frac{1}{2}$ defects relegated to the boundary. This insensitivity to topological constraints distinguishes active from passive liquid crystals.

\begin{figure}[h]
\includegraphics[width=\columnwidth]{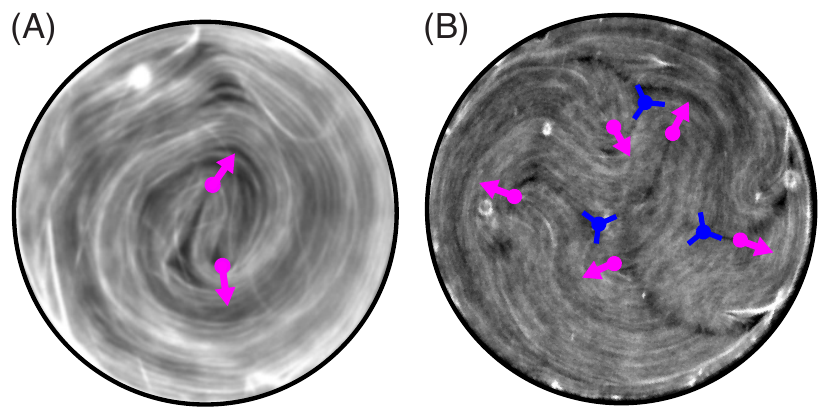}
\caption{(A and B) Fluorescence microscopy images of the kinesin-microtubule system described in the text that motivate our theoretical investigation, with microtubules fluorescently labeled and confined to SU8 holes with radii of (A) $50 \mu$m (see movie S1) and (B) $250 \mu$m (movie S2). Defects are labeled with magenta arrows ($+\frac{1}{2}$) and blue points with three spokes ($-\frac{1}{2}$).}\label{expvthry}
\end{figure}

\begin{figure}
\includegraphics[width=\columnwidth]{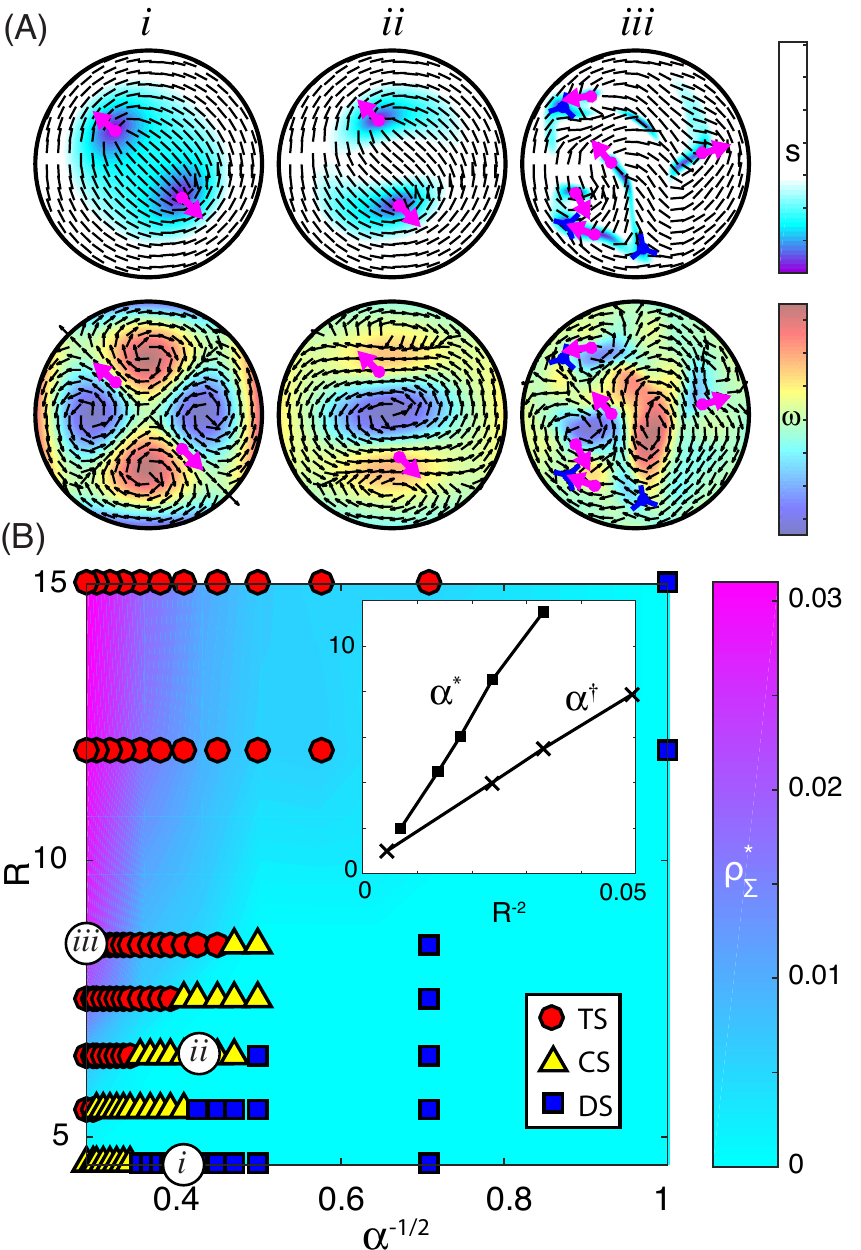}
\caption{(A) Director and order (top), and vorticity and streamlines (bottom) corresponding to the three dynamical steady states observed in the FEM simulations: (i)  dipolar state (DS, movie S3), (ii) circulating state (CS, movie S4) and (iii) turbulent state (TS, movie S5). Results are shown for parallel anchoring. (B)  Time-averaged, excess  defect density  $\rhoD^*=\left(\ND-2\right)/\pi R^2$ (where $\ND$ is the total number of $\pm\frac{1}{2}$ defects) as a function of disk radius $R$ and active length scale $\alpha^{-1/2}$\cite{Giomi2015, Doostmohammadi2016, Hemingway2016a}. We subtract 2 to offset by the number of topologically required $+\frac{1}{2}$ defects. The three cases shown in (A) are indicated with white circles. The inset of (B) plots the threshold activities for the transitions from DS to CS ($\alpha^{\dagger}$) and CS to TS ($\alpha^*$) as a function of $R^{-2}$.}
\label{phasediagram}
\end{figure}

\section{\label{sec:level1} Model} Our study is motivated by experiments on a widely studied model active nematic system comprising microtubule bundles driven by ATP-powered kinesin motor proteins \cite{DeCamp2015,Henkin2014,Sanchez2012,Sanchez2011}. Recently this system has been studied under hydrodynamic and topological confinement by placing the suspension in microfoabricated SU8 wells that are $O\left(100\mu \text{m}\right)$ in diameter and enforce parallel anchoring of microtubules. Examples of configurations observed in these experiments are shown in \fig\ref{expvthry} and the corresponding dynamics are shown in movies S1 and S2; a future work will explore the experimental system in more detail.

As a minimal representation of this system, we use a single-fluid continuum model whose state is described by the dimensionless nematic order tensor $\mathbf{Q}=s\rho\left[\mathbf{n}\otimes\mathbf{n}-\left(1/2\right)\mathbf{I}\right]$ and fluid flow field $\mathbf{u}$. $\mathbf{Q}$ describes both the local orientation $\mathbf{n}$ and degree of order $s$ of the nematic and is scaled by the nematic density $\rho$ such that ($\rho s = \sqrt{2 \Tr \mathbf{Q}^2}$). The coupled dynamics are given by
 \begin{equation}
 \partial_{t}\mathbf{Q}+\nabla\cdot\left(\mathbf{u}Q\right)=\left(\mathbf{Q}\mathbf{\Omega}-\mathbf{\Omega} \mathbf{Q}\right)
+\lambda\mathbf{E}^{\tau}+\gamma^{-1}\mathbf{H}
 \end{equation}

\begin{figure*}
\includegraphics[width=\textwidth]{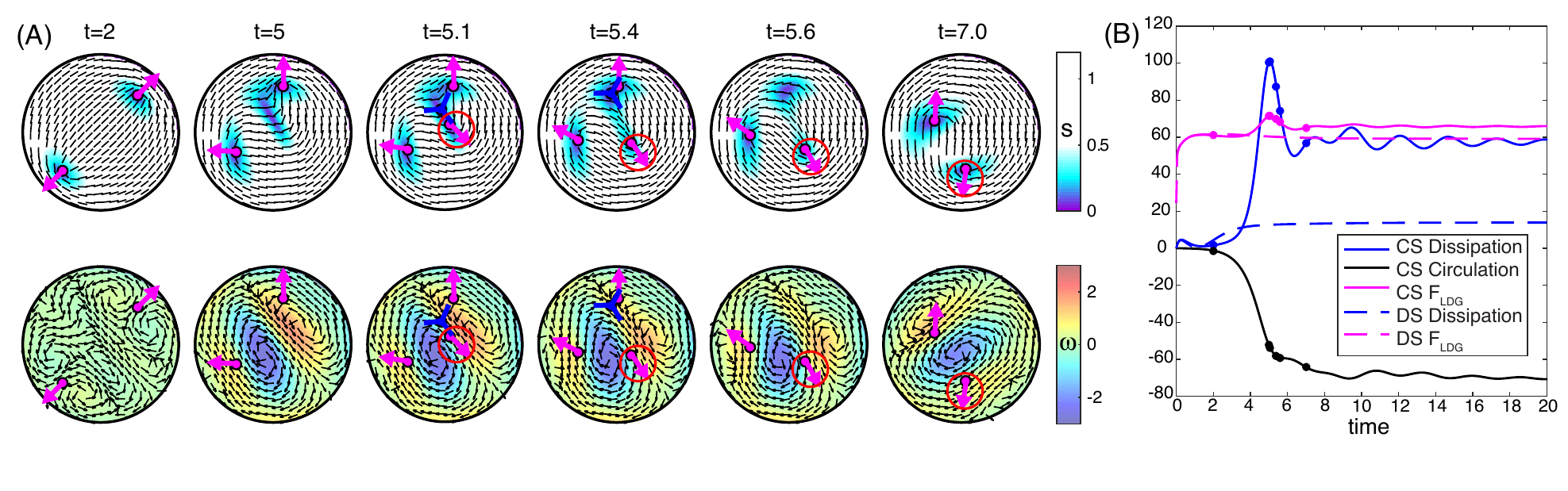}
\caption{(A) Simulation results showing the director and degree of order (top) and flow field and vorticity (bottom) during the transition from the DS to the CS; $\alpha=5$ and $R=6.5$, the same as \fig 2.A.ii (movie S4). The stills show the process in six steps: (1) The dipole configuration similar to that shown in \fig 2.A.i. (2) Two $+\frac{1}{2}$ defects with a nascent $\pm\frac{1}{2}$ pair (purple region of disorder). (3) The ejection of a $+\frac{1}{2}$ defect (circled in red to distinguish it from the other $+\frac{1}{2}$ defects). (4 \& 5) The annihilation of one of the original DS $+\frac{1}{2}$ defects. (6) The beginning of the CS. (B) Total system viscous dissipation (blue), $\mathcal{F}_\text{LDG}$ (magenta) and circulation (black) as a function of time during the transition from the DS to the CS shown in \fig ~\ref{fig:DStoCStransition} ($R=6.5$ and $\alpha=5$). The dashed lines show results when we impose left-right symmetry, suppressing the CS and thus forcing the system to remain in the DS.}
\label{fig:DStoCStransition}
\end{figure*}

Kinematic terms and free-energy relaxation both contribute to the dynamics of $\mathbf{Q}$. The kinematic terms depend on the local fluid flow velocity and gradients, with $\Omega_{ij}=\frac{1}{2}\left(\partial_{i}u_{j}-\partial_{j}u_{i}\right)$ as the antisymmetric vorticity tensor and $E_{ij}=\frac{1}{2}\left(\partial_{i}u_{j}+\partial_{j}u_{i}\right)$ as the symmetric strain rate tensor. The operation $A^{\mathcal{T}}$ denotes the traceless version of a second order tensor $A_{ij}^{\tau}=A_{ij}-\frac{1}{2}\delta_{ij}A_{kk}$. The relaxational terms are proportional to variations of the system free energy, with $H_{ij}=-\delta \mathcal{F}/ \delta Q_{ij}$ and $\gamma^{-1}$ as the dissipation rate. The total free energy of the system is given by $\mathcal{F}=\mathcal{F}_{\text{LDG}}+\mathcal{F}_{\text{ND}}$ where the first term
\begin{multline}
\mathcal{F}_\text{LDG}=\int_{\Omega} d^2\mathbf{r}\left\{C\left(-\frac{\beta_1}{2}\Tr\left(\mathbf{Q}^2\right)+\frac{\beta_2}{4}\Tr\left(\mathbf{Q}^2\right)^2\right)\right.\\
\left.+\frac{1}{2}L_1\left|\nabla\mathbf{Q}\right|^2+\frac{1}{2}L_2\left(\nabla\cdot\mathbf{Q}\right)\cdot\left(\nabla\cdot\mathbf{Q}\right)\right\},
\label{eq:fLDG}
\end{multline}
is the bulk Landau-deGennes free energy  \cite{DeGennes1995}. The dimensionless functions $\beta_1\left(\rho\right)=\rho-1$ and $\beta_2\left(\rho\right)=\left(\rho+1\right)/\rho^{2}$ control the transition from an isotropic fluid ($\rho<1$) to a nematic phase ($\rho>1$); in this work we set $\rho=1.6$ to focus on the nematic phase far away from the phase transition. The second free energy term
\begin{align}
\mathcal{F}_\text{ND}=\oint_{\partial\Omega} d\mathbf{r}\frac{1}{2}E_{A}\Tr\left(\left(\mathbf{Q}-\mathbf{W}\right)^{2}\right)
\label{eq:fND},
\end{align}
is the Nobili-Durand boundary anchoring energy \cite{Nobili1992} with the director and order along the boundary given by the tensor $\mathbf{W}$; a similar form of the anchoring energy has been used previously in the study of active nematic suspensions \cite{Giomi2014c}. For example, parallel anchoring on a circular boundary with boundary tangent $\mathbf{t}\left(\theta\right)=\{-\sin{\left(\theta\right)},\cos{\left(\theta\right)}\}$ gives $\mathbf{W}=s^{*}\rho \left[\mathbf{t}\otimes\mathbf{t}-\left(1/2\right)\mathbf{I}\right]$ where $s^*=\sqrt{2}$ is the degree of order associated with a nematic in the limit $\rho\rightarrow\infty$ \cite{Putzig2015}. The gradient descent dynamics are therefore given by
\begin{multline}
 \gamma^{-1}H_{ij}=\Dr\left(\beta_1-\beta_2 Q_{kl}Q_{lk}\right)Q_{ij}+2\DE\partial_{k}\partial_{k}Q_{ij}
\\-\DA\left(Q_{ij}-W_{ij}\right)|_{\partial\Omega},
\label{eq:Q1}
\end{multline}
where $\DE=\left(L_1+L_2\right)/2\gamma$, $\DA=E_A/\gamma$ and $\Dr=C/\gamma$.

Momentum conservation in the Stokes limit along with incompressibility constraint $\nabla\cdot\mathbf{u}=0$ governs the fluid flow
\begin{align}
\eta\nabla^{2}\mathbf{u}-\nabla P-\alpha \nabla\cdot\mathbf{Q}+\nabla\cdot\sigma_\text{p}=0
\label{eq:Stokes}
\end{align}
with pressure $P$, dynamic viscosity $\eta$, strength of activity $\alpha$, and passive elastic stress tensor $\sigma_\text{p}=-\lambda s \mathbf{H}+ \mathbf{QH}-\mathbf{HQ}$. The active stress, $-\alpha\mathbf{Q}$ corresponds to an extensile dipole force density and is the leading order active term that can arise from a nematic fluid; gradients in the director and order, $\nabla\cdot\mathbf{Q}$, impart force into the fluid \cite{Simha2002, Marchetti2013, Thampi2014}. Including an active stress term of this form is motivated by the observed extensile nature of the microtubule system, \fig\ref{expvthry} \cite{Sanchez2011, DeCamp2015}.

We non-dimensionalize the system using the time scale $T=\Dr^{-1}$ and length scale $L=\sqrt{\DE/\Dr}$ (equivalently, $L=\sqrt{\left(L_1+L_2\right)/2C}$) and introducing dimensionless  operators ($\bar{\partial}_{t}=\partial_{t}/\Dr$, $\bar{\partial}_{i}=\partial_{i}/\sqrt{\DE/\Dr}$). This gives the dimensionless system
\begin{align*}
\bar{\partial}_{t}\mathbf{Q}+\bar{\nabla}\cdot\left(\bar{\mathbf{u}}Q\right)=\left(\mathbf{Q}\bar{\mathbf{\Omega}}-\bar{\mathbf{\Omega}} \mathbf{Q}\right)
+\bar{\lambda}\bar{\mathbf{E}}^{\tau}+\bar{\mathbf{H}},
\end{align*}
\begin{align*}
\bar{H}_{ij}=\left(\beta_1-\beta_2 Q_{kl}Q_{lk}\right)Q_{ij}+2\bar{\partial}_{k}\bar{\partial}_{k}Q_{ij}
\\-\bar{D}_A\left(Q_{ij}-W_{ij}\right)|_{\partial\Omega},
\end{align*}
\begin{align}
\bar{\nabla}^{2}\mathbf{\bar{u}}-\bar{\nabla}\bar{P}-\bar{\alpha}\bar{\nabla}\cdot\mathbf{Q}+\bar{\nabla}\cdot\bar{\sigma}_\text{p}=0
\label{eq:fullsys}
\end{align}
with variables ($\bar{u}=u\sqrt{\Dr\DE}$ and $\bar{P}=P/\Dr\eta$ and parameters $\bar{D}_A=\DA/\Dr$, $\bar{\lambda}=\lambda/\Dr$, $\bar{\alpha}=\alpha/\eta \Dr$. Outside of this section, reference to the active stress strength will always refer to the dimensionless quantity $\bar{\alpha}$.

We solve eqns. \eqref{eq:fullsys} in a circular domain of radius $R$. We assume no-slip boundary conditions on the fluid flow ($\bar{\mathbf{u}}|_{\partial\Omega}=0$) and arbitrarily fix pressure to $\bar{P}=0$ at a point along the container wall. We intialize the flow field at rest, $\bar{\mathbf{u}}=0$, and the director is set to a small random perturbation of 5\% around a uniform field. We assume that the active stress and viscous dissipation dominate the force balance and therefore neglect passive elastic stresses (by setting $\bar{\sigma}_\text{p}=0)$. As we will show, defect densities scale as expected from theories which include these additional forces, and our phase diagram (\fig\ref{phasediagram}) is similar to that from a recent numerical study on confined active nematics that includes these additional terms \cite{Doostmohammadi2017}.

To integrate the equations of motion, we used the finite element analysis software COMSOL by inputting the equations directly using the weak form. We removed second order derivatives using integration by parts, creating natural boundary conditions $\bar{\mathbf{u}}|_{\partial\Omega}=0$ and $\nabla\mathbf{Q}|_{\partial\Omega}=0$. The former was explictly overwritten with the no-slip condition, while the boundary energy term $\propto \bar{D}_\text{A}$ contributes to the latter. We used quadratic elements for $\mathbf{Q}$ and $\bar{\mathbf{u}}$, and linear elements for the pressure/continuity equation \cite{Donea2003}. The element size was  $\Delta x\sim 0.1$. The largest system we considered ($R=15$) produced a system with $\sim 10^5$ DOF, which was completed in a few hours on a desktop computer. We tracked defects in the simulation results using the same software developed to study experimental systems in DeCamp \etal \cite{DeCamp2015}.

\section{\label{sec:level1} Results and Discussion}
\subsection{\label{sec:level1} Parallel Anchoring}
Previous works have yet to study the boundary conditions that most closely represent the experimental system: no-slip hydrodynamic boundary conditions ($\mathbf{u}|_{\partial\Omega}=0$) and parallel anchoring of the nematic $\mathbf{w}=\{-\sin{\left(\theta\right)},\cos{\left(\theta\right)}\}$ such that the net topological charge is +1. Here, we begin by focusing on these boundary conditions, which represent a topologically incommensurate confinement in that a defect-free nematic cannot be formed. We consider a range of domain sizes ($R=4.5-15$) and active stress strengths ($\alpha=0-12$). Experimentally these parameters are controlled by, respectively, varying the microfluidic well radius, and motor protein concentration. In principle the nematogen density $\rho$ can also be varied, but this is harder to control experimentally and so we leave it fixed at $\rho=1.6$. Moreover, for the model considered here which does not lead to concentration gradients, increasing density maps to increasing $\alpha$ (to leading order for $\rho>1$). Finally, we fix the boundary relaxation term to $\DA=3$ for all simulations and assume our material to be flow aligning with $\lambda=1$.

We observe three dynamical steady states as confinement and activity are varied:
\textit{(i)} At high confinement (small $R$) and low $\alpha$, we observe a stationary state that is topologically identical to the equilibrium configuration. For parallel anchoring, this consists of two static $+\frac{1}{2}$ defects located at antipodal positions, and directed radially outward (\fig\ref{phasediagram}A.i); we refer to this as the dipolar state (DS). Although the director is static, the active stress generates a quadrapolar flow with four equally-sized vortices. In this regime, the director relaxation dominates over flow-alignment.

\textit{(ii)} As the activity level or container size are increased past threshold values, the system transitions to a state in which the two $+\frac{1}{2}$ defects circulate in closed orbits (\fig\ref{phasediagram}A.ii); we refer to this configuration as the circulating state (CS). Importantly, while the rearrangement of the director configuration and defect orientations between the DS and CS appears small, the scale and structure of vorticity has changed dramatically. The four equally sized vortices of the DS are replaced by two smaller vortices with the same sense of rotation, and a single large vortex with opposite sign. Vortex coarsening and circulation have been observed in systems such as the microtubule-kinesin system (\fig \ref{expvthry}), swimming organisms \cite{Wensink2012,Wioland2013,Lushi2014,Wioland2016}, crawling cells \cite{Duclos2014,Segerer2015, Duclos2016} and previous numerical studies of confined active nematics with different boundary conditions \cite{Woodhouse2012, Gao2017, Shendruk2017}. We show below that this state is highly robust to boundary conditions. Consistent with earlier findings using natural boundary conditions on $\mathbf{Q}$\cite{Woodhouse2012}, the threshold activity $\alpha^{\dagger}$ for transitioning from stationary to persistent defect circulation depends on domain size according to $\alpha^{\dagger}\sim R^{-2}$ (inset of \fig\ref{phasediagram}.B).

Additionally, we identified striking, symmetry-breaking dynamics during the development of the CS, \fig\ref{fig:DStoCStransition}A and movie S4. By starting with initial conditions close to the DS configuration, we observed how the director and flow fields evolve during the dynamical transition into the CS. The resulting trajectories show that the system momentarily creates a region of disorder in the form of a new $\pm\frac{1}{2}$ pair. The $-\frac{1}{2}$ defect created during this event rapidly annihilates with one of the original dipolar $+\frac{1}{2}$ defects, creating the co-rotating defect configuration. Based on observations of simulation trajectories, we conjecture that the DS$\rightarrow$CS transition requires creating a region of local disorder, such as an additional $\pm\frac{1}{2}$defect pair. While the simulation trajectories show that there is sufficient active energy to produce excess defects within the DS, the system no longer produces excess defects once the transition to the topologically equivalent CS is complete. While the mechanism for stability is unexplored, this observation suggests that the hydrodynamics of the CS inhibit the continued formation of defects.

To quantify the differences between the CS and DS, we plot the free energy $\mathcal{F_{\text{LDG}}}$, total system dissipation, and the total circulation $\oint_{\partial\Omega}\mathbf{u}\cdot\mathbf{e}_{\theta}\text{d}\mathbf{x}$ in \fig\ref{fig:DStoCStransition}B. To compare the CS and DS states at the same parameter values, we present results from an additional set of simulations that enforced left-right symmetry (thus suppressing the CS and retaining the DS, dashed lines in \fig\ref{fig:DStoCStransition}B). We see that $\mathcal{F^{\text{CS}}_{\text{LDG}}}>\mathcal{F^{\text{DS}}_{\text{LDG}}}$. While the director fields for the CS and DS are similar, the CS is, in fact, more deformed.  Additionally, during the transition ($t=5-6$) the free energy of the CS exceeds its steady value when the additional pair is created, indicating the presence of an activation barrier for the process. The dissipation in the CS is also greater than in the DS. This can be understood by noting that the force imparted by the nematic into the fluid $\propto\nabla\cdot\mathbf{Q}$; thus, to leading order more distortions lead to more energy imparted into the flow, which subsequently must be dissipated.

\textit{(iii)} Above a threshold radius/activity, defects proliferate and the system transitions into a turbulent state (TS) that qualitatively resembles the behavior of an unconfined active nematic \cite{Giomi2015, Putzig2015, Oza2015b} (\fig\ref{phasediagram}A.iii). We define $\ND$ as the average number of both $\pm\frac{1}{2}$ defects; the excess defect density is then $\rhoD^*=\left(\ND-2\right)/\pi R^2$. Beyond the transition, $\rhoD^*$ scales linearly with the offset activity $\alpha-\alpha^*$ as it does in the bulk \cite{Giomi2015}. We therefore define $\alpha^*$ as the point at which bulk defect density scale begins; $\alpha^*\propto R^{-2}$ (see inset of \fig\ref{phasediagram}.B). The phase diagram presented is consistent with earlier findings \cite{Doostmohammadi2017, Gao2017}.

\begin{figure}
\includegraphics[width=3.0in]{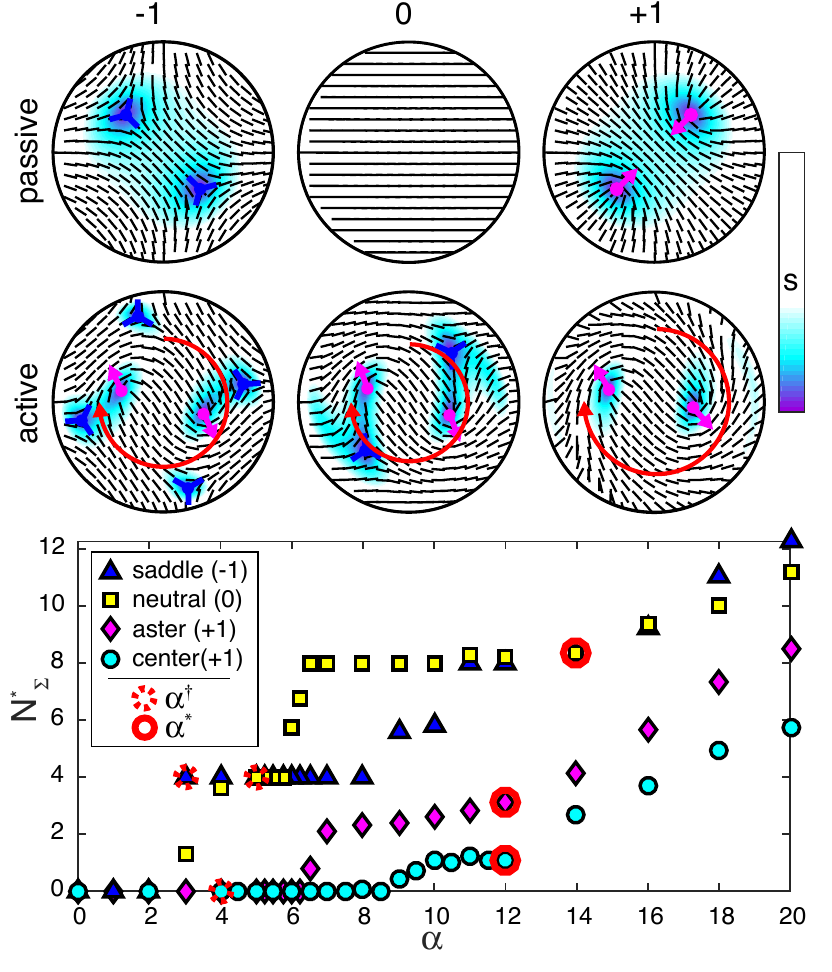}
\caption{Behaviors in three different container topologies: (from left to right) saddle (-1), neutral (0), aster (+1). (A) (top) Equilibrium configuration for $\alpha=0$, and (bottom) circulating states observed for $\alpha=5$. Animations of the circulating states are provided in movies 6-9.  (B) Excess defect number $N^*_{\Sigma}$ as a function of activity for each topology with $\alpha^{\dagger,*}$ labelled. $R=6.5$ for all cases.
\label{fig:topologicalBC}}
\end{figure}

\begin{figure}[h]
\includegraphics[width=\columnwidth]{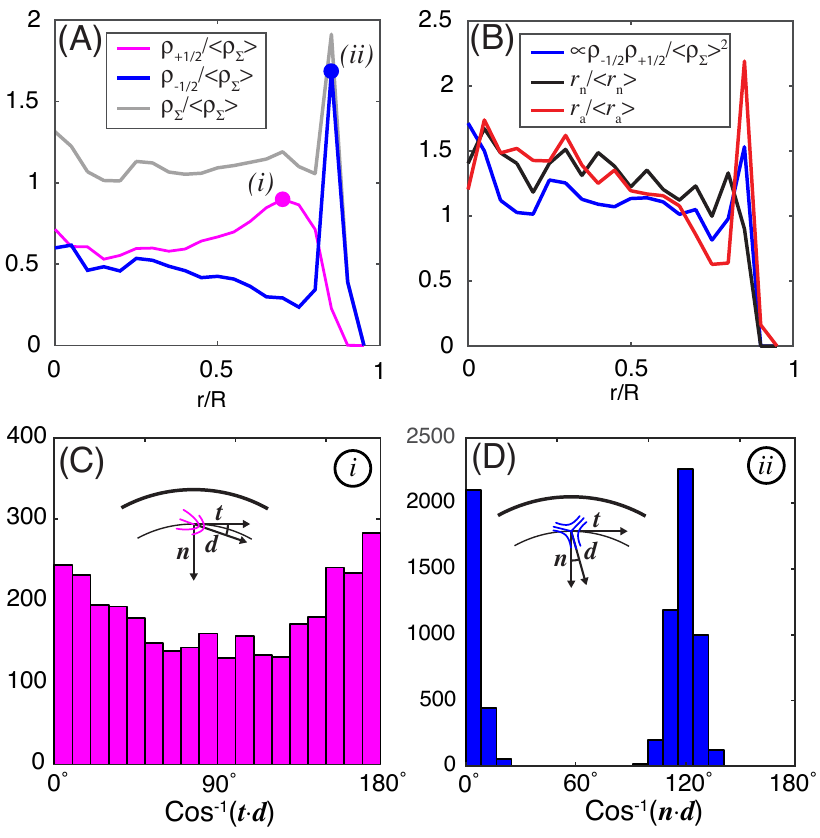}
\caption{(A) Spatial distributions of $+\frac{1}{2}$ (magenta), $-\frac{1}{2}$ (blue), and all (gray) defects normalized by the total, average defect concentration for the same conditions in the TS ($\alpha=12$, $R=15$). (B) Spatial distribution of defect nucleation rates (black) and  annihilation rates (red) normalized by the system average. To illustrate the extent to which annihilation rates follow the law of mass action, the product of the spatial distributions of $\pm\frac{1}{2}$ defects (scaled by a factor of 4 to make the trend the same order of magnitude as the creation/destruction rates) is also plotted (blue). (C) and (D) plot the normalized orientation distributions for each defect type at the radial points corresponding to their respective maxima (\emph{i}) and (\emph{ii}).
}
\label{fig:TurbDefectDist}
\end{figure}

\subsection{\label{sec:level1} Other topologies}
We now explore the effects of container-imposed topological constraints on system behaviors by considering anchoring conditions that favor aster, neutral, or saddle director configurations at equilibrium. Respectively, these conditions correspond to net topological charges of $-1$, 0, and $+1$, and we construct them by specifying an anchoring director $\mathbf{w}$ along the boundary given in cartesian vectors by $\{\cos{\theta},-\sin{\theta}\}$, $\{1,0\}$, and $\{\cos{\theta},\sin{\theta}\}$. We note that independent control of the boundary conditions for the flow field and director can only be achieved with ``wet" nematic models such as used here. Overdamped or ``dry" models simplify the governing equations by assuming $\mathbf{u}\propto\nabla\cdot\mathbf{Q}$ \cite{Putzig2014, Putzig2015, Oza2015b, Oza2016b, DeCamp2015, Ramaswamy2010, Kumar2014}; however, this assumption precludes prescribing the no-slip boundary condition for arbitrary boundary geometry and topology. Previous theoretical works have explored the transition between these limits in boundary-less systems\cite{Thampi2014a, Doostmohammadi2016}; here, we focus on the wet limit for simplicity.
Because our results with parallel anchoring show that increasing activity can be mapped to decreasing radius, we fix the container radius at $R=6.5$ and vary activity $\alpha$.

For all topologies, we observe the same three classes of steady states described for parallel anchoring (\fig\ref{fig:topologicalBC}). At low activity (\ie high confinement) we observe the topologically minimal state for each container, consistent with its equilibrium configuration. Above a threshold activity $\alpha^\dagger$, whose value depends only weakly on topology, the system transitions to a circulating state with two co-rotating $+\frac{1}{2}$ defects and sufficient $-\frac{1}{2}$ defects to fulfill the topological constraint. While the $-\frac{1}{2}$ defects contribute to the flow, their influence decays rapidly in space \cite{Giomi2014} and they tend to reside along the boundary. The system-sized vortex structure is therefore preserved across topologies.
Above a higher threshold activity $\alpha^*$, which also depends only weakly on topology, the system transitions to the TS, with the defect number proportional to $\alpha$ as discussed above for parallel anchoring.
The most significant difference between topologies occurs between the onset of the CS and the transition to the TS. The neutral topology admits a second sub-turbulent state with two additional $+\frac{1}{2}$ defects with more complex, but still regular defect trajectories (movie S8); the dynamics strongly resemble the ``dancing defect" state observed in topologically neutral channels \cite{Shendruk2017}. This additional state suggests the possibility of finely-tuned non-trivial active states in the range $\alpha^{\dagger}<\alpha<\alpha^*$ for these and other topologies not considered here.

Deep in the turbulent regime ($\alpha=12$ and $R=15$) we find that defects exhibit non-trivial spatial distributions and orientations near the boundary. \fig\ref{fig:TurbDefectDist}A shows the time-averaged spatial distributions of defects for the parallel anchoring container; both defect types accumulate near the boundary, but at different radial positions respectively labeled by \emph{i} and \emph{ii}. The $-\frac{1}{2}$ defects are located close to the wall; the radial position of this maxima is anchoring condition dependent but scales like the active length scale $\alpha^{-1/2}$. In contrast, $+\frac{1}{2}$ defects are displaced further from the wall, toward the center. Because of these displaced and non-uniform distributions, the net topological charge of the container (+1 in this case) is distributed unevenly throughout the system. There is an `interior region' where there are equal populations of $\pm\frac{1}{2}$; since it is topologically neutral we consider the interior region to be bulk-like. This is surrounded by a `topological boundary layer' containing the displaced peaks of $\pm\frac{1}{2}$, and a net topological charge of $+1$.

In \fig\ref{fig:TurbDefectDist}B we examine the spatial distribution of annihilation and nucleation and consider whether they are the cause of the of the spatially nonuniform defect densitities, \fig\ref{fig:TurbDefectDist}A. We see that nucleation rates are nearly uniform throughout the domain, indicating no spatially preferred sites of defect generation. Annihilation rates are peaked in the boundary layer; however, \fig\ref{fig:TurbDefectDist}B shows that the distribution of annihilation rates is roughly proportional to the product of defect densities, $\rhoprod(r)=\rho_+(r) \rho_{-}(r)$. This suggests that defect annihilation simply follows the law of mass action for a bimolecular reaction; thus, the spatial dependence of defect annihilation rates is a consequence (rather than a cause) of the nonuniform defect density. Taken together, these trends suggest that the spatially nonuniform defect distributions arise because the inner and outer boundary regions act as attractors for +$\frac{1}{2}$ and -$\frac{1}{2}$ defects.

Defect dynamics are complex because defects with different charges have qualitatively different hydrodynamics \cite{Giomi2014}; this leads us to hypothesize that the different locations of stability for $\pm\frac{1}{2}$ defects reflect differences in anisotropic hydrodynamic wall interactions of the defect species.  In particular, $-\frac{1}{2}$ defects are stable in orientations for which their active flow pushes them toward the wall, while $+\frac{1}{2}$ defects are unstable in such orientations. In support of this hypothesis, \fig ~\ref{fig:TurbDefectDist}.C and D. show the orientational distributions of $\pm\frac{1}{2}$ defects, measured at their respective locations of maximal density. We see that $-\frac{1}{2}$ defects have a strong tendency to orient with one of their ``points'' facing inward, normal to the wall. Near the wall, the three-fold symmetry of the flow is broken, leading to a net active flow which drives the defect further into the wall. In contrast, $+\frac{1}{2}$ defects tend to orient tangentially to the wall, such that their active flow drives them to process around the container. In the final section, we perform an axillary analysis to confirm the stability of these orientations for both defects in the absence of other forces.

\begin{figure}[h]
\includegraphics[width=\columnwidth]{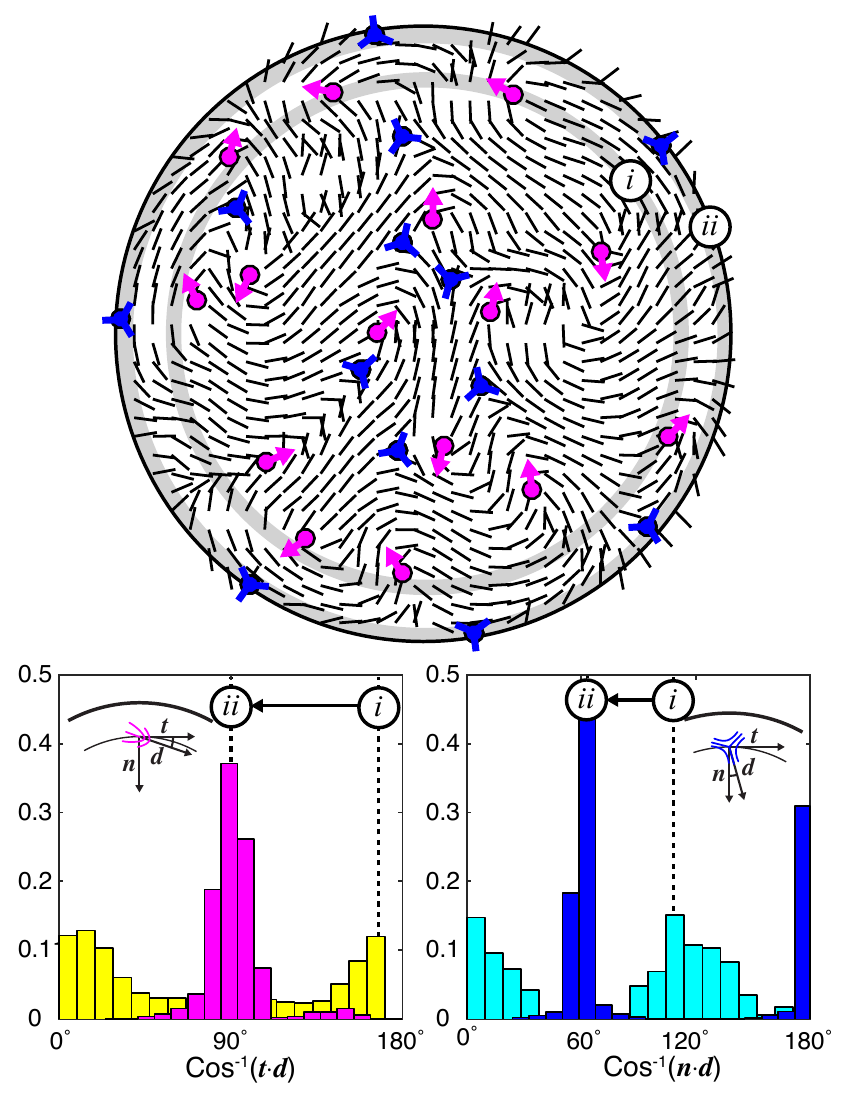}
\caption{(Top) Director field and defects in the TS ($\alpha=12$, $R=15$) for aster (+1) boundary condition (distributions for additional topologies are shown in the appendix \fig\ref{fig:topBCorientation}). Labeled annuli have width $\Delta r=0.83$ with inner radii $R-3\Delta r$ and $R- \Delta r$, and are used to calculate orientation probability distributions of $+/-\frac{1}{2}$ (bottom left/right) at regions dominated by hydrodynamic wall interactions (\emph{i}) and anchoring energy (\emph{ii}), respectively.}
\label{fig:top_orient_dist}
\end{figure}

This behavior persists regardless of topology. \fig\ref{fig:top_orient_dist} compares defect orientions at two annuli for the aster (+1) boundary condition and once again finds a hydrodynamic region (a) with peaks identical to \fig\ref{fig:TurbDefectDist}. Only close to the wall (b) are orientations perturbed by anchoring, \fig\ref{fig:top_orient_dist}. Additional topologies are presented in \fig\ref{fig:topBCorientation}.

\subsection{\label{sec:level1}Hydrodynamic Stability of Defects Near Walls}

To assess the orientational hydrodynamic stability of $\pm\frac{1}{2}$ defects near a wall, we solve the Stoke's equation (Eqn. \ref{eq:Stokes2}), in the presence of an imposed force distribution that represents an idealized defect. We perform the calculation separately for the two defect charges; in each case the force is imposed over a discrete region. We then find the net vorticity on the defect resulting from the imposed force, $\bar{\omega}=\int_{0}^{r_0} r\omega dr$.  This approach only evaluates the orientational stability of a defect with respect to hydrodynamic forces.

\begin{figure}
\includegraphics[scale=1.0]{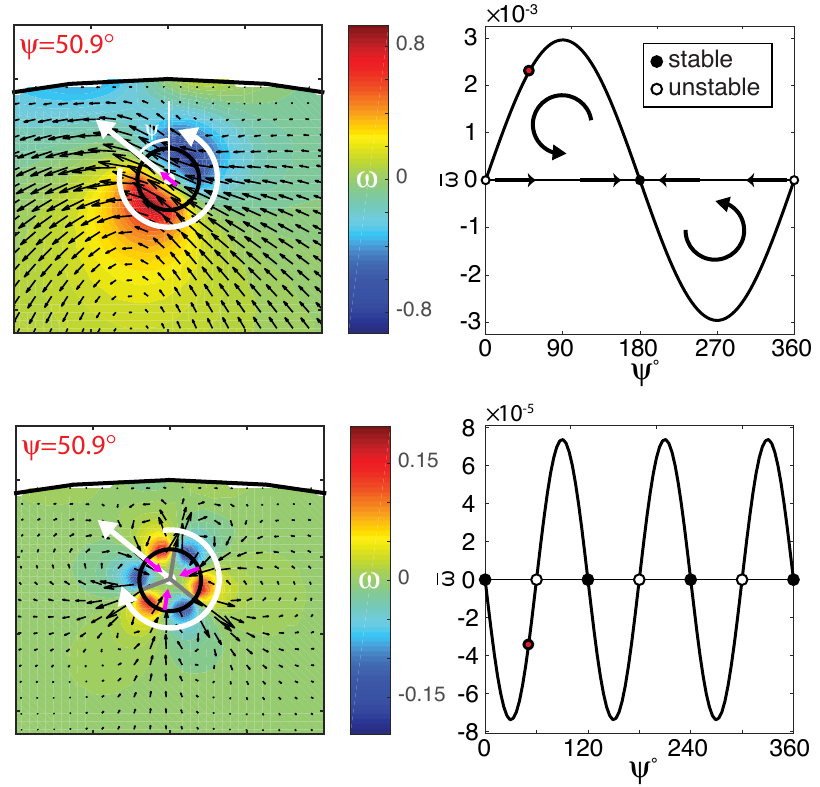}
\caption{(Left) Flow field and vorticity for $+(-)\frac{1}{2}$ defects near a wall at angle $\psi$. The area over which the force is imposed is circumscribed by a black circle in each plot. (Right) The total vorticity integrated over the defect area as a function of orientation $\psi$, revealing both fixed points (roots) and stability (slope). Filled (open) circles denote stable (unstable) stationary defect orientations. The red circle in each plot corresponds to the configuration in the left column.}
\label{fig:DefectStability}
\end{figure}

Details of the calculation are as follows. The Stokes equation is written as:
\begin{align}
\nabla^{2}\mathbf{u}-\nabla P+\Theta\left(r-r_0\right)f_0\mathbf{v}_{\pm\frac{1}{2}}\left(\psi\right)=0
\label{eq:Stokes2},
\end{align}
with the third term approximating a defect with regions of uniform force density. For more detail on the flow created by single defects we refer the reader to \fig 2 of Giomi \etal\cite{Giomi2014}. For a $+\frac{1}{2}$ defect, the force is defined as a disk of radius $r_0$ (black circle) (where $\Theta$ is the Heaviside step function) with a force per unit area of uniform magnitude $f_0$, in the direction $\mathbf{v}_{+\frac{1}{2}}\left(\psi\right)$ (where $\psi$ is the polar angle defined in \fig\ref{fig:DefectStability}):
\begin{align}
\mathbf{v_{+\frac{1}{2}}}=
\left \{
  \begin{tabular}{ccc}
  $-\sin{\psi}$ \\
  $\cos{\psi}$
  \end{tabular}
\right \}
\label{eq:forceplus}
\end{align}

Similarly, a $-\frac{1}{2}$ defect is represented as a sum of three discrete regions, each with an inward facing force oriented $120^{\circ}$ with respect to its neighbors:
\begin{align}
\mathbf{v_{-\frac{1}{2}}}=\sum_{i=1}^{3}{\Theta_{i}\left(\theta,\psi\right)}
\left \{
  \begin{tabular}{ccc}
  $\sin{\left(\psi+\delta_i\right)}$ \\
  $-\cos{\left(\psi+\delta_i\right)}$
  \end{tabular}
\right \}
\end{align}
where the step-functions define the three angular sectors around the defect center $\Theta_{i}\left(\theta,\psi\right)=\Theta\left(\theta-\left(\frac{\pi}{6}+\delta_i\right)-\psi\right)-\Theta\left(\theta-\left(\frac{5\pi}{6}+\delta_i\right)-\psi\right)$, and the angle $\delta_i=\left(i-1\right)\frac{2\pi}{3}$ defines the angle shift between sectors.

\fig\ref{fig:DefectStability} shows an example of the flow and vorticity fields (left column) created by each of the defects near a wall; the right column plots the resulting net vorticity on the defect region as a function of orientation angle $\psi$. Each defect has multiple stationary orientations (corresponding to zero net vorticity), but only some of these configurations are stable. For example, the $+\frac{1}{2}$ defect can be oriented normal to the wall directed inward $\psi=0^{\circ}$ or facing away $\psi=180^{\circ}$, but since $\frac{\partial\bar{\omega}}{\partial\psi}|_{\psi=0} > 0$, perturbations will result in a rotation away from $\psi=0$. Since the stable orientation for $+\frac{1}{2}$ defects faces away from the boundary, they will tend to both reorient and propel away from boundaries. In contrast, for $-\frac{1}{2}$ defects $\frac{\partial\bar{\omega}}{\partial\psi}|_{\psi=0,\frac{2\pi}{3},\frac{4\pi}{3}} < 0$, so vorticity will restore the defect to orientations $\psi=0$, $\frac{2\pi}{3}$ and $\frac{4\pi}{3}$. Unlike the stable configuration of $+\frac{1}{2}$, in these configurations, the flow continues to drive the defect into the wall.

The consequences of hydrodynamic wall interactions with different anchoring conditions are shown in \fig\ref{fig:TurbDefectDist} and \fig\ref{fig:top_orient_dist}; an exhaustive accounting of the behaviour of all topologies explored is given in \fig~\ref{fig:topBCorientation}. In all cases, the annular region closest to the wall is dominated by $\mathcal{F}_\text{LDG}$. Throughout the rest of the domain, hydrodynamics dominate and system behaviors are topology-independent.

\section{\label{sec:level1} Conclusion}

In a confined passive liquid crystal, the director field is globally determined by the topology imposed by chemistry and geometry of the boundary. Our results show that in an active liquid crystal, defect hydrodynamics relegate any net topological charge required by the container's geometry and boundary conditions to a small layer along the boundary; this creates a topologically neutral, bulk-like interior. The overall spatiotemporal dynamics are therefore insensitive to boundary conditions on the director field. Remarkably, this insensitivity persists even under sufficiently high confinement to establish the minimal motile configuration, the circulating state, whose flow consists of the coarsest possible vortex geometry. In all topologies that we explored, the active flows created by the two co-rotating $+\frac{1}{2}$ defects dominate the flow. While the inability of anchoring to affect system behaviors suggests that passive liquid crystal control strategies cannot be directly applied to active systems, the persistence of the circulating state in different container topologies (including those without explicit topological constraints \cite{Woodhouse2012, Gao2017}) suggests robustness that could be leveraged to design microfluidic systems containing active nematics.

\section{\label{sec:level1} Acknowledgements}

This work was supported by the NSF MRSEC-1420382 (AB,OA,SF,AB,MFH); NSF DMR-1149266 (AB)); NSF DMREF-1534890 (MMN, BL); and DOE DE-SC0010432TDD (OA)).



\begin{thebibliography}{10}

\bibitem{Bisoyi2011}
H.~K. Bisoyi and S.~Kumar, ``{Liquid-crystal nanoscience: an emerging avenue of
  soft self-assembly},'' {\em Chem. Soc. Rev.}, vol.~40, no.~1, pp.~306--319,
  2011.

\bibitem{Senyuk2013}
B.~Senyuk, Q.~Liu, S.~He, R.~D. Kamien, R.~B. Kusner, T.~C. Lubensky, and I.~I.
  Smalyukh, ``{Topological colloids.},'' {\em Nature}, vol.~493, no.~7431,
  pp.~200--5, 2013.

\bibitem{Luo2016}
Y.~Luo, F.~Serra, D.~A. Beller, M.~A. Gharbi, N.~Li, S.~Yang, R.~D. Kamien, and
  K.~J. Stebe, ``{Around the corner: Colloidal assembly and wiring in groovy
  nematic cells},'' {\em Physical Review E}, vol.~93, no.~3, pp.~1--8, 2016.

\bibitem{Peng2016}
C.~Peng, T.~Turiv, Y.~Guo, S.~V. Shiyanovskii, Q.-H. Wei, and O.~D.
  Lavrentovich, ``{Control of colloidal placement by modulated molecular
  orientation in nematic cells},'' {\em Science Advances}, vol.~2, no.~9,
  pp.~1--9, 2016.

\bibitem{Simha2002}
A.~Simha and S.~Ramaswamy, ``{Hydrodynamic fluctuations and instabilities in
  ordered suspensions of self-propelled particles},'' {\em Physical review
  letters}, vol.~89, pp.~1--4, jul 2002.

\bibitem{Saintillan2013a}
D.~Saintillan and M.~J. Shelley, ``{Active suspensions and their nonlinear
  models},'' {\em Comptes Rendus Physique}, vol.~14, no.~6, pp.~497--517, 2013.

\bibitem{Marchetti2013}
M.~C. Marchetti, J.~F. Joanny, S.~Ramaswamy, T.~B. Liverpool, J.~Prost, M.~Rao,
  and R.~A. Simha, ``{Hydrodynamics of soft active matter},'' {\em Reviews of
  Modern Physics}, vol.~85, no.~3, pp.~1143--1189, 2013.

\bibitem{Voituriez2005}
R.~Voituriez, J.-F. Joanny, and J.~Prost, ``{Spontaneous flow transition in
  active polar gels},'' {\em Europhysics Letters}, vol.~70, p.~7, may 2005.

\bibitem{Edwards2009}
S.~a. Edwards and J.~M. Yeomans, ``{Spontaneous flow states in active nematics:
  a unified picture},'' {\em Europhysics Letters}, vol.~85, p.~6, jan 2008.

\bibitem{Giomi2011}
L.~Giomi and M.~C. Marchetti, ``{Polar patterns in active fluids},'' {\em Soft
  Matter}, vol.~8, no.~1, pp.~129--139, 2012.

\bibitem{Giomi2012}
L.~Giomi, L.~Mahadevan, B.~Chakraborty, and M.~F. Hagan, ``{Banding,
  excitability and chaos in active nematic suspensions},'' {\em Nonlinearity},
  vol.~25, no.~8, pp.~2245--2269, 2012.

\bibitem{Woodhouse2012}
F.~G. Woodhouse and R.~E. Goldstein, ``{Spontaneous circulation of confined
  active suspensions},'' {\em Physical Review Letters}, vol.~109, pp.~1--5,
  oct 2012.

\bibitem{Wioland2013}
H.~Wioland, F.~G. Woodhouse, J.~Dunkel, J.~O. Kessler, and R.~E. Goldstein,
  ``{Confinement stabilizes a bacterial suspension into a spiral vortex},''
  {\em Physical Review Letters}, vol.~110, no.~26, pp.~1--5, 2013.

\bibitem{Ravnik2013}
M.~Ravnik and J.~M. Yeomans, ``{Confined active nematic flow in cylindrical
  capillaries},'' {\em Physical Review Letters}, vol.~110, pp.~1--5, jan 2013.

\bibitem{Doostmohammadi2016}
A.~Doostmohammadi, M.~F. Adamer, S.~P. Thampi, and J.~M. Yeomans,
  ``{Stabilization of active matter by flow-vortex lattices and defect
  ordering.},'' {\em Nature communications}, vol.~7, pp.~1--9, 2016.

\bibitem{Wioland2016}
H.~Wioland, E.~Lushi, and R.~E. Goldstein, ``{Directed collective motion of
  bacteria under channel confinement},'' {\em New Journal of Physics}, vol.~18,
  no.~7, pp.~27--30, 2016.

\bibitem{Shendruk2017}
T.~N. Shendruk, A.~Doostmohammadi, K.~Thijssen, J.~M. Yeomans, J.~M. Yeomans,
  J.~M. Yeomans, R.~A. Simha, F.~Mecarini, F.~D. Angelis, E.~D. Fabrizio, and
  S.~Eaton, ``{Dancing disclinations in confined active nematics},'' {\em Soft
  Matter}, vol.~13, no.~21, pp.~3853--3862, 2017.

\bibitem{Gao2017}
T.~Gao, M.~D. Betterton, A.-S. Jhang, and M.~J. Shelley, ``{Analytical
  structure, dynamics, and coarse-graining of a kinetic model of an active
  fluid},'' {\em arXiv.org}, pp.~1--33, 2017.

\bibitem{Wu2017}
K.-T. Wu, J.~B. Hishamunda, D.~T.~N. Chen, S.~J. DeCamp, Y.-W. Chang,
  A.~Fern{\'{a}}ndez-Nieves, S.~Fraden, and Z.~Dogic, ``{Transition from
  turbulent to coherent flows in confined three-dimensional active fluids},''
  {\em Science}, vol.~355, no.~6331, p.~eaal1979, 2017.

\bibitem{Doostmohammadi2017}
T.~N. Shendruk, K.~Thijssen, J.~M. Yeomans, and A.~Doostmohammadi, ``{Onset of
  meso-scale turbulence in active nematics},'' {\em Nature Communications},
  vol.~8, no.~May, pp.~1--7, 2017.

\bibitem{Sokolov2010}
A.~Sokolov, M.~M. Apodaca, B.~A. Grzybowski, and I.~S. Aranson, ``{Swimming
  bacteria power microscopic gears},'' {\em Proceedings of the National Academy
  of Sciences}, vol.~107, pp.~969--974, jan 2010.

\bibitem{Thampi2016}
S.~P. Thampi, A.~Doostmohammadi, T.~N. Shendruk, R.~Golestanian, and J.~M.
  Yeomans, ``{Active micromachines: Microfluidics powered by mesoscale
  turbulence.},'' {\em Science advances}, vol.~2, no.~7, p.~e1501854, 2016.

\bibitem{Zhang2016}
R.~Zhang, Y.~Zhou, M.~Rahimi, and J.~J. de~Pablo, ``{Dynamic structure of
  active nematic shells},'' {\em Nature Communications}, vol.~7, p.~13483,
  2016.

\bibitem{Giomi2015}
L.~Giomi, ``{Geometry and topology of Turbulence in active nematics},'' {\em
  Physical Review X}, vol.~5, no.~3, pp.~1--10, 2015.

\bibitem{Hemingway2016a}
E.~J. Hemingway, P.~Mishra, M.~C. Marchetti, and S.~M. Fielding, ``{Correlation
  lengths in hydrodynamic models of active nematics},'' {\em Soft Matter},
  vol.~12, pp.~7943--7952, 2016.

\bibitem{DeCamp2015}
S.~J. Decamp, G.~S. Redner, A.~Baskaran, M.~F. Hagan, and Z.~Dogic,
  ``{Orientational order of motile defects in active nematics},'' {\em
  arXiv.org}, vol.~32, pp.~1--13, aug 2015.

\bibitem{Henkin2014}
G.~Henkin, S.~J. DeCamp, D.~T.~N. Chen, T.~Sanchez, and Z.~Dogic, ``{Tunable
  dynamics of microtubule-based active isotropic gels},'' {\em Philosophical
  Transactions of the Royal Society A: Mathematical, Physical and Engineering
  Sciences}, vol.~372, no.~2029, pp.~20140142--20140142, 2014.

\bibitem{Sanchez2012}
T.~Sanchez, D.~T.~N. Chen, S.~J. Decamp, M.~Heymann, and Z.~Dogic,
  ``{Spontaneous motion in hierarchically assembled active matter},'' {\em
  Nature}, vol.~491, pp.~1--5, nov 2012.

\bibitem{Sanchez2011}
T.~Sanchez, D.~Welch, D.~Nicastro, and Z.~Dogic, ``{Cilia-Like Beating of
  Active Microtubule Bundles},'' {\em Science}, vol.~333, no.~6041, pp.~456
  --459, 2011.

\bibitem{DeGennes1995}
P.~G. de~Gennes and J.~Prost, {\em {The Physics of Liquid Crystals}}.
\newblock Oxford University Press, 1995.

\bibitem{Nobili1992}
M.~Nobili and G.~Durand, ``{Disorientation-induced disordering at a
  nematic-liquid-crystal-solid interface},'' {\em Physical Review A}, vol.~46,
  pp.~R6174--R6177, nov 1992.

\bibitem{Giomi2014c}
L.~Giomi and A.~Desimone, ``{Spontaneous division and motility in active
  nematic droplets},'' {\em Physical Review Letters}, vol.~112, no.~14,
  pp.~1--5, 2014.

\bibitem{Putzig2015}
E.~Putzig, G.~S. Redner, A.~Baskaran, and A.~Baskaran, ``{Instabilities,
  defects, and defect ordering in an overdamped active nematic},'' {\em Soft
  Matter}, vol.~12, no.~17, pp.~1--5, 2015.

\bibitem{Thampi2014}
S.~P. Thampi, R.~Golestanian, and J.~M. Yeomans, ``{Vorticity, defects and
  correlations in active turbulence.},'' {\em Philosophical transactions.
  Series A, Mathematical, physical, and engineering sciences}, vol.~372,
  no.~2029, pp.~431--434, 2014.

\bibitem{Donea2003}
J.~Donea and A.~Huerta, {\em {Adaptive Finite Element Method for Thermal Flow
  Problems}}.
\newblock John Wiley {\&} Sons, Ltd, 1994.

\bibitem{Wensink2012}
H.~H. Wensink, J.~M. Yeomans, R.~E. Goldstein, J.~Dunkel, S.~Heidenreich,
  K.~Drescher, R.~E. Goldstein, H.~Lowen, and J.~M. Yeomans, ``{Meso-scale
  turbulence in living fluids},'' {\em Proceedings of the National Academy of
  Sciences}, vol.~109, pp.~14308--14313, sep 2012.

\bibitem{Lushi2014}
E.~Lushi, H.~Wioland, and R.~E. Goldstein, ``{Fluid flows created by swimming
  bacteria drive self-organization in confined suspensions.},'' {\em
  Proceedings of the National Academy of Sciences of the United States of
  America}, vol.~111, no.~27, pp.~9733--9738, 2014.

\bibitem{Duclos2014}
G.~Duclos, S.~Garcia, H.~G. Yevick, and P.~Silberzan, ``{Perfect nematic order
  in confined monolayers of spindle-shaped cells},'' {\em Soft Matter},
  vol.~10, no.~14, pp.~2346--2353, 2014.

\bibitem{Segerer2015}
F.~J. Segerer, F.~Th{\"{u}}roff, A.~{Piera Alberola}, E.~Frey, and J.~O.
  R{\"{a}}dler, ``{Emergence and persistence of collective cell migration on
  small circular micropatterns},'' {\em Physical Review Letters}, vol.~114,
  no.~22, pp.~1--5, 2015.

\bibitem{Duclos2016}
G.~Duclos, C.~Erlenk{\"{a}}mper, J.-F. Joanny, and P.~Silberzan, ``{Topological
  defects in confined populations of spindle-shaped cells},'' {\em Nature
  Physics}, vol.~13, no.~1, pp.~1--6, 2016.

\bibitem{Oza2015b}
A.~U. Oza and J.~Dunkel, ``{Antipolar ordering of topological defects in active
  liquid crystals},'' {\em New Journal of Physics}, vol.~18, no.~9, pp.~1--8,
  2016.

\bibitem{Putzig2014}
E.~Putzig and A.~Baskaran, ``{Phase separation and emergent structures in an
  active nematic fluid},'' {\em Physical Review E - Statistical, Nonlinear, and
  Soft Matter Physics}, vol.~90, pp.~1--9, oct 2014.

\bibitem{Oza2016b}
A.~U. Oza, S.~Heidenreich, and J.~Dunkel, ``{Generalized Swift-Hohenberg models
  for dense active suspensions},'' {\em European Physical Journal E}, vol.~39,
  no.~10, p.~97, 2016.

\bibitem{Ramaswamy2010}
S.~Ramaswamy, ``{The mechanics and statistics of active matter},'' {\em Annual
  Review of Condensed Matter Physics}, vol.~1, pp.~323--345, apr 2010.

\bibitem{Kumar2014}
N.~Kumar, H.~Soni, S.~Ramaswamy, and A.~K. Sood, ``{Flocking at a distance in
  active granular matter},'' {\em Nature Communications}, vol.~5, p.~4688,
  2014.

\bibitem{Thampi2014a}
S.~P. Thampi, R.~Golestanian, and J.~M. Yeomans, ``{Active nematic materials
  with substrate friction},'' {\em Physical Review E - Statistical, Nonlinear,
  and Soft Matter Physics}, vol.~90, no.~6, pp.~1--6, 2014.

\bibitem{Giomi2014}
L.~Giomi, M.~J. Bowick, P.~Mishra, R.~Sknepnek, and M.~C. Marchetti, ``{Defect
  dynamics in active nematics},'' {\em arXiv.org}, vol.~372, p.~16, oct 2014.

\end{thebibliography}

\renewcommand{\thefigure}{A\arabic{figure}}
\setcounter{figure}{0}
\section*{Appendix}

\begin{figure*}
\includegraphics[width=\textwidth]{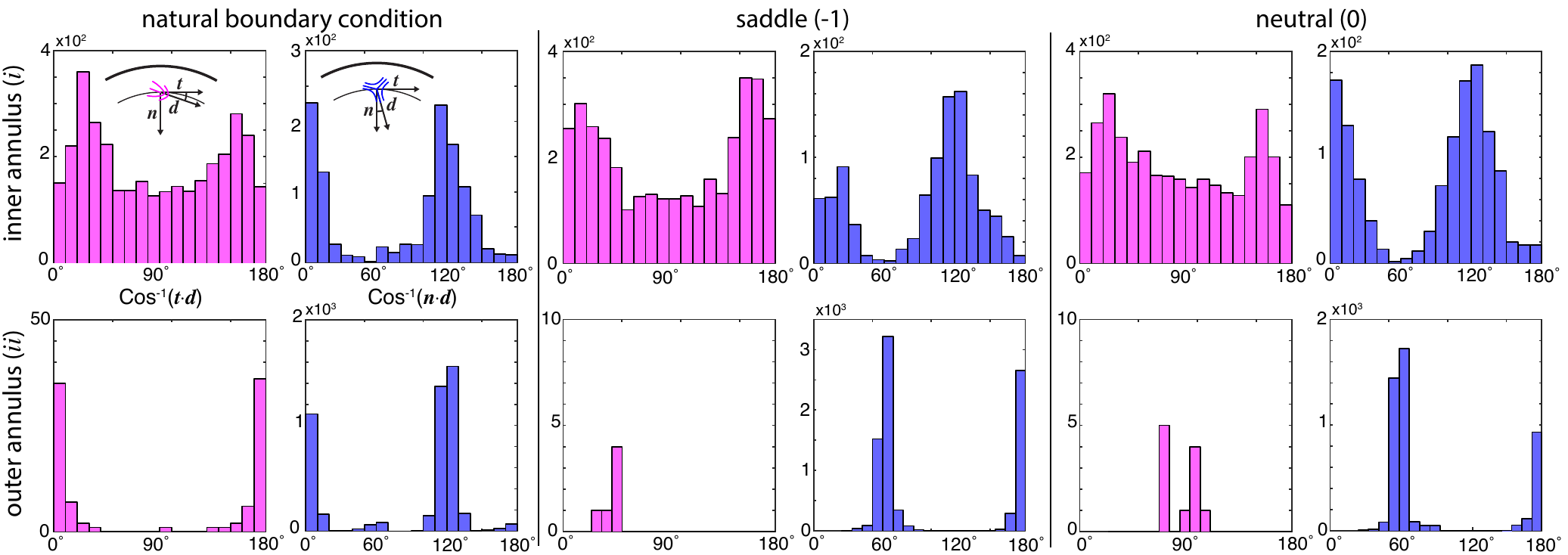}
\caption{Orientation distributions for $\pm\frac{1}{2}$ defects (magenta/blue) in the turbulent regimes ($\alpha=12$, $R=15$) for 3 different topological boundary conditions; from left to right: natural boundary condition $\mathbf{n}\cdot\mathbf{Q}|_{\partial\Omega}=0$ (no topological preference), saddle (-1), neutral (0). Rows are different radial bins (see \fig\ref{fig:top_orient_dist}). In bin (\emph{i}), defect orientations are dominated by hydrodynamic wall effects, and are therefore similar for all anchoring conditions. In bin (\emph{ii}), the importance of the anchoring condition near the wall is revealed by differences in orientation distributions for different topological boundary conditions.}
\label{fig:topBCorientation}
\end{figure*}

\textbf{Movie S1:}
An experiment showing the microtubule-kinesin suspension \cite{DeCamp2015} on an oil/water interface confined to a SU8 well, $R=100\mu\text{m}$. Defects are tracked using the algorithm described in \cite{DeCamp2015} and labeled: $+\frac{1}{2}$ (magenta arrows) and $-\frac{1}{2}$ (blue points with three-prongs). The circulating state observed is evocative of simulation results (movie S4) but is periodically interrupted by nucleation of a new $\pm\frac{1}{2}$ defect pairs at the boundary that completely invade the domain. These new defects eventually anneal, re-establishing the circulating state. The movie is played at 15 fps, the time elapsed between frames is $\Delta t=$ 1s.

\textbf{Movie S2:}
An experiment showing the microtubule-kinesin suspension \cite{DeCamp2015} on an oil/water interface confined to a SU8 well, $R=250\mu\text{m}$, with defects tracked \cite{DeCamp2015} and labeled: $+\frac{1}{2}$ (magenta arrows) and $-\frac{1}{2}$ (blue points with three-prongs). The phenomenology of the system closely resembles the turbulent state described in the main text, movie S5. The movie plays at 10 fps, and the time elapsed between frames is $\Delta t=$ 2s.

\textbf{Movie S3:}
Dynamics of the dipolar state (DS). The movie shows the time evolution of the director and order (left), and flow field and vorticity (right) from an FEM simulation. Defects are labeled in both: $+\frac{1}{2}$ (magenta arrows) and $-\frac{1}{2}$ (blue points with three-prongs). Parameters are: $\alpha=5.5$, $R=4.5$, parallel anchoring and no-slip boundary conditions, non-dimensional time between frames of $\Delta t=0.04$, and the movie plays at 30 fps.

\textbf{Movie S4:}
Transition to the circulating state (CS), and dynamics within the CS. The movie shows the director and order (left), and flow field and vorticity (right) from an FEM simulation. Defects are labeled in both: $+\frac{1}{2}$ (magenta arrows) and $-\frac{1}{2}$ (blue points with three-prongs). Parameters are: $\alpha=5$, $R=6.5$, $\Delta t=0.02$, played at 30 fps. Stills from the transition from initial condition (which corresponds to the DS) to the circulating state CS are shown in \fig\ref{fig:DStoCStransition}.

\textbf{Movie S5:}
Dynamics of the turbulent state (TS). The movie shows the director and order (left), and flow field and vorticity (right), with defects labeled in both: $+\frac{1}{2}$ (magenta arrows) and $-\frac{1}{2}$ (blue points with three-prongs). Parameters are: $\alpha=12$, $R=8.5$, $\Delta t=0.02$, played at 30 fps. We consider the regular defect trajectories and annihilation/nucleation patterns from the first third of the video as transient behavior, since once turbulent dynamics set in these patterns do not return. The calculations for defect density and nucleation/annihilation rate use dynamics only from the latter $2/3$ of trajectories.

\textbf{Movie S6:}
Development of the CS in a FEM simulation with the saddle (-1) topological boundary condition, aand $\alpha=5$, $R=6.5$, and $\Delta t=0.02$. Left: Director (lines) and order (color). Right: flow field (quiver) and vorticity (color). Defects are labeled in both: $+\frac{1}{2}$ (magenta arrows) and $-\frac{1}{2}$ (blue points with three-prongs).

\textbf{Movie S7:}
Development of the CS in a FEM simulation with the neutral (0) topological boundary condition, and  $\alpha=5$, $R=6.5$, and $\Delta t=0.02$. Left: Director (lines) and order (color). Right: flow field (quiver) and vorticity (color). Defects are labeled in both: $+\frac{1}{2}$ (magenta arrows) and $-\frac{1}{2}$ (blue points with three-prongs).

\textbf{Movie S8:}
The development of a higher order circulating state during a FEM simulation with the neutral (0) topology, at a higher activity than movie S7, $\alpha=8$, and $R=6.5$ and $\Delta t=0.02$. Left: Director (lines) and order (color). Right: flow field (quiver) and vorticity (color). Defects are labeled in both: $+\frac{1}{2}$ (magenta arrows) and $-\frac{1}{2}$ (blue points with three-prongs). The steady state dynamics consist of 4 $+\frac{1}{2}$ defects that dart across the center of the domain in an alternating fashion; their motion periodically changes the sign of circulation.

\textbf{Movie S9:}
Development of the CS in a FEM simulation with the aster (+1) topological boundary condition, and $\alpha=5$, $R=6.5$, and $\Delta t=0.02$. Left: Director (lines) and order (color). Right: flow field (quiver) and vorticity (color). Defects are labeled in both: $+\frac{1}{2}$ (magenta arrows) and $-\frac{1}{2}$ (blue points with three-prongs).

\end{document}